\documentclass[manuscript]{acmart}
\setcopyright{none}

\AtBeginDocument{%
  \providecommand\BibTeX{{%
    \normalfont B\kern-0.5em{\scshape i\kern-0.25em b}\kern-0.8em\TeX}}}

\copyrightyear{2024}
\acmYear{2024}

\acmConference[FAccT '24]{}{Non-Archival
  }{2024}
\usepackage[subtle]{savetrees}
\begin{document}

\title{The Legal Duty to Search for Less Discriminatory Algorithms}

\author{Emily Black}
\authornote{Both authors contributed equally to this research.}
\authornote{This work is an abridged version of a law review article entitled \emph{Less Discriminatory Algorithms} by the same authors, which is forthcoming in the Georgetown Law Journal, October 2024, Vol. 113~\cite{black2023less}. Interested readers can find the piece at \url{https://papers.ssrn.com/sol3/papers.cfm?abstract_id=4590481}. This non-archival version of the paper was presented at FAccT 2024.}
\email{eblack@barnard.edu}
\affiliation{%
  \institution{Barnard College, Columbia University}
  \country{USA}
  }
\author{Logan Koepke}
\authornotemark[1]
\email{logan@upturn.org}
\affiliation{%
  \institution{Upturn}
  \country{USA}
}

\author{Pauline Kim}

\email{kim@wustl.edu}
\affiliation{%
  \institution{Washington University in St. Louis}
  \country{USA}
}

\author{Solon Barocas}

\affiliation{%
  \institution{Microsoft Research}
  \country{USA}
}

\author{Mingwei Hsu}

\affiliation{%
  \institution{Upturn}
  \country{USA}
}

\renewcommand{\shortauthors}{Black and Koepke, et al.}

\begin{abstract}
Work in computer science has established that, contrary to conventional wisdom, for a given prediction problem there are almost always multiple possible models with equivalent performance---a phenomenon often termed model multiplicity. Critically, different models of equivalent performance can produce different predictions for the same individual, and, in aggregate, exhibit different levels of impacts across demographic groups. As a result, when an algorithmic system displays a disparate impact, model multiplicity suggests that developers may be able to discover an alternative model that performs equally well, but has less discriminatory impact. Indeed, the promise of model multiplicity is that an equally accurate, but less discriminatory algorithm (LDA) almost always exists. But without dedicated exploration, it is unlikely developers will discover potential LDAs. Model multiplicity and the availability of less discriminatory algorithms have significant ramifications for the legal response to discriminatory algorithms, in particular for disparate impact doctrine, which has long taken into account the availability of alternatives with less disparate effect when assessing liability. A close reading of legal authorities over the decades reveals that the law has on numerous occasions recognized that the existence of a less discriminatory alternative is sometimes relevant to a defendant’s burden of justification at the second step of disparate impact analysis. Indeed, under disparate impact doctrine, it makes little sense to say that a given algorithmic system used by an employer, creditor, or housing provider is either “justified” or “necessary” if an equally accurate model that exhibits less disparate effect is available and possible to discover with reasonable effort. As a result, we argue that the law should place a duty of a reasonable search for LDAs on entities that develop and deploy predictive models in covered civil rights domains.   
\end{abstract}

\maketitle

\section{Introduction}
For years, advocates have expressed concerns that reliance on algorithmic systems in areas covered by civil rights laws---such as tenant screening systems, employment assessment and hiring technologies, and credit underwriting models---will further discrimination and exclusion of historically marginalized groups. Over the past decade, a vast debate has emerged on how law, policy, and regulation should respond to discriminatory algorithms.

An often unspoken premise throughout this debate is that for any given prediction problem, a single ``correct” model exists. For example, when a bank seeks to predict default by borrowers, it is often assumed that a single ``correct” model exists that best meets that goal, and that any deviation from this unique solution would entail a loss of performance. The implication is that pursuing goals like minimizing discrimination will unavoidably involve a tradeoff with accuracy. But the assumptions that a unique solution exists and that a fairness-accuracy tradeoff is inevitable are descriptively inaccurate. Recent work in computer science has established that there are almost always multiple possible models with equivalent accuracy for a given optimization problem---a phenomenon that has been named ``the Rashomon effect,''~\cite{breiman2001statistical} ``predictive multiplicity''~\cite{marx2019}, and ``model multiplicity''~\cite{blackmodel2022}, among others~\cite{black2021leave,d2020underspecification}. 

Multiplicitous models perform a given prediction task equally well, but may differ in other ways---from the features they use to make their predictions, to the way they combine features to make their predictions, to the way their predictions are robust to changing circumstances. 
Recent work demonstrates that, both in theory~\cite{blackmodel2022,semenova2022existence} and in practice~\cite{rodolfa2021empirical,pmlr-v139-coston21a,d2020underspecification,auerbach2024testing}, a range of models can perform similarly for a given prediction task.
Critically, these equally performant models can have different levels of disparate impact on historically disadvantaged groups~\cite{rodolfa2021empirical,pmlr-v139-coston21a, auerbach2024testing}.
As a result, when an algorithmic system displays a disparate impact, model multiplicity suggests that other models that perform equally well, but have less discriminatory effect, exist. 
In this paper, we refer to such models as less discriminatory algorithms (LDAs).

The existence of LDAs has significant ramifications for American civil rights law, particularly the disparate impact doctrine. 
The disparate impact doctrine aims to prevent discrimination through decision-making systems that are facially neutral---i.e., do not take a protected attribute or related proxy into account in its decisions---yet have discriminatory effects, such as credit underwriting models that disproportionately approve loans for white applicants compared with Black applicants, or hiring systems that select a higher proportion of men than women. Existing federal laws in employment, housing, and credit prohibit practices that have an unjustified disparate impact. At a high-level, the disparate impact doctrine is similar across these areas and operates through a three-step process. First, a plaintiff must establish a prima facie case that a business' decision-making practice leads to disparate effects. Next, the defendant (the business with the practice in question) can avoid liability by demonstrating a legitimate business justification for the practice. Even if it does so, however, it can still face liability if the plaintiff offers a \emph{less discriminatory alternative}, which advances the same business needs, but with less disparate impact. 

As we detail in Section~\ref{sec:legal}, the definitions surrounding the requirements for a less discriminatory alternative are ill-defined legally. However, under the most restrictive approach, a less discriminatory alternative must be both ``equally effective'' to the original decision-making system, and must reduce harmful impact on a disadvantaged group.
In the algorithmic setting, one can translate ``equally effective'' into ``equally performant on whatever metric a given model is trained to optimize for'' (e.g., accuracy in predicting the likelihood that a borrower will default). And in keeping with legal standards from civil rights laws, we can define an algorithm as less discriminatory compared to another when it displays less disparity in selection rates across demographic groups. In this way, model multiplicity speaks directly to the legal concept of a less discriminatory alternative because it suggests that there will almost always be less discriminatory \emph{algorithms} (LDAs) that could serve as an alternative to a baseline model that has disparate effects. 


Given this observation, we argue that in the case of algorithmic decision-making systems, it makes little sense to require a plaintiff to search for LDAs. Because model multiplicity implies that LDAs exist when a model displays disparate effects, the relevant question should not be ``can the plaintiff offer an alternative{?}" but ``what reasonable steps has the defendant taken to discover LDAs{?}'' Compared to defendants, plaintiffs are poorly positioned to discover LDAs: they may not have sufficient resources, technical knowledge, or access to relevant data and modeling pipelines. Developers at businesses covered by the doctrine, meanwhile, are best positioned to search for LDAs because developing a model inherently involves constant exploration and testing of alternatives. 
Requiring entities to also test for disparate impact and compare model disparities throughout the model development process is fairly easy and is, by itself, not burdensome. Several methods for doing so are publicly available~\cite{black2023toward}.

Thus, our core argument is that entities that use algorithmic systems in domains like housing, employment, and credit should have a duty to search for and implement LDAs before they can deploy a system with disparate effects. Without such a duty, developers are likely to be singularly focused on their chosen performance metric and will fail to identify ways to achieve the same goals with less discriminatory impact. By placing the burden to search for a less discriminatory alternative on the better-resourced and better-positioned actor, the doctrine will be more effective in preventing algorithmic harms.
Our paper proceeds as follows: in Section~\ref{sec:model_mult}, we provide a brief review of model multiplicity. In Section~\ref{sec:legal}, we provide a summary of the legal background and history of the disparate impact doctrine, showing that previous authorities sometimes considered the availability of alternatives when deciding whether a defendant met its burden of justification. In Section ~\ref{sec:mm_and_law}, we connect model multiplicity to discrimination law, and describe the policy and regulatory interventions that would recognize this duty to search.  In Section~\ref{sec:upstart}, we describe a case study involving a real-world example of a search for an LDA, and in Sections~\ref{sec:duty} we describe the requirements and practical implementation of this duty to search in detail. Finally, in Section~\ref{sec:caveats}, we discuss the caveats and limitations to our proposal. 





\section{Model Multiplicity}
\label{sec:model_mult}
\begin{figure*}[b]
    \centering
    \includegraphics[width=0.9\textwidth]{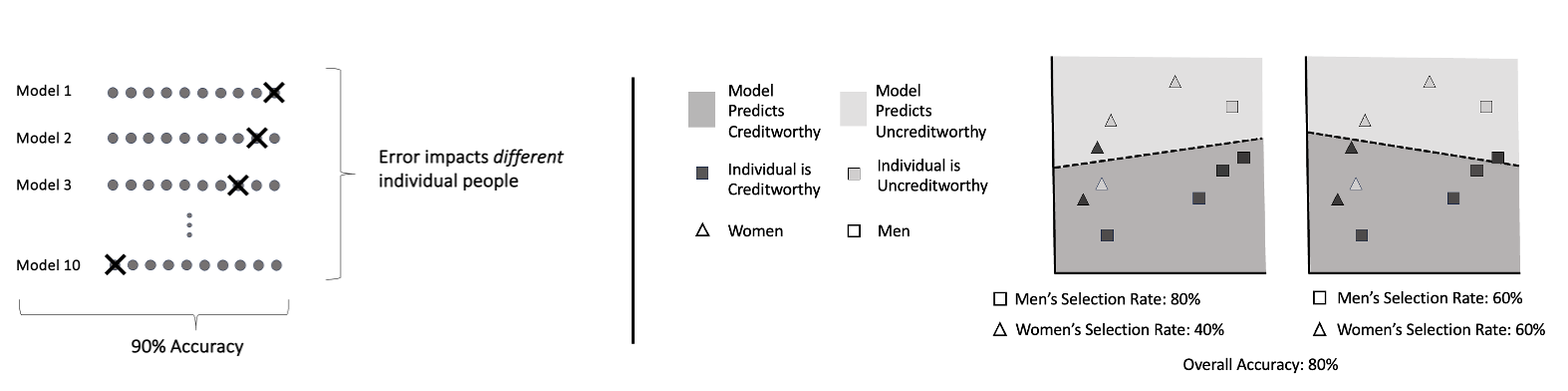}
    
    \vspace{-1.5em}
    \caption{Left: An illustration of how ten different models can exhibit the same accuracy while giving different
individual predictions on a hypothetical group of ten people. Right: An example of two multiplicitous models: they display equal accuracy (80\% over all people), yet make different individual predictions, leading to a difference in discriminatory behavior. The graph to the left has a steep difference in selection rate between men and women, whereas the graph to the right does not. The darker region of the graph refer to places where the model predicts an individual to be creditworthy, and darker points correspond to individuals who are indeed creditworthy. The lighter region of the graph refers to areas where the model predicts an individual to be uncreditworthy, and lighter points correspond to individuals who are indeed uncreditworthy. Triangular points refer to women, and square points refer to men.}
    \label{fig:equal_acc}
\end{figure*}
In this section, we provide a brief review of model multiplicity, define what we refer to as less discriminatory algorithms (LDAs), and describe how to find LDAs in practice.
\subsection{Model Multiplicity and LDAs}

A range of work describes how 
there often exist many equally performant models for a given prediction problem~\cite{breiman2001statistical,pmlr-v139-coston21a,marx2019,black2021leave,paes2023inevitability,semenova2019study,dong2019variable,rodolfa2021empirical,blackmodel2022}. 
While many terms have been used to describe related phenomena, we use the term model multiplicity~\cite{blackmodel2022}. We define two models to have equivalent performance if the metrics they were trained to optimize are within some contextually-relevant threshold $\epsilon$ on a given test set. For the sake of simplicity we will use accuracy as a stand-in for any more specific performance metric, but note that our arguments apply to context-relevant performance metrics beyond accuracy.

We do not go into the technical background of this phenomenon given its coverage in the literature, but point to Black et al.~\cite{blackmodel2022} for an overview. At a basic level, multiple models with the same performance exist because, for any given error rate, there are different ways to distribute accurate predictions over a population (see Figure ~\ref{fig:equal_acc}). 
So long as a model's accuracy is below the optimal Bayes error, model multiplicity guarantees that another model with similar overall accuracy exists that would generate predictions differently~\cite{blackmodel2022}. When taken in aggregate, these differences in predictions may produce different levels of disparities across groups. Thus, unless a company has fortuitously discovered the model with minimum possible disparity among all equivalent models, in many cases the phenomenon of model multiplicity guarantees that there exists a model with indistinguishable accuracy but less disparate impact---i.e., a less discriminatory algorithm (LDA). The notion of discrimination can be defined in a variety of ways, but in this paper, we draw on the legal doctrine of disparate impact, which focuses on disparities in selection rates that disadvantage historically marginalized groups.

Finding a \emph{specific} equally accurate model---corresponding to a particular re-drawing of a model’s decision boundary---is difficult through the typical model development process. At the same time, finding equally accurate models in general is easy, as they are common. Indeed, research has shown that models with equivalent accuracy which differ in other behavior (including disparate impact) can be easily discovered in practice, and occur naturally throughout the model development process~\cite{rodolfa2021empirical,pmlr-v139-coston21a,auerbach2024testing}. As a result, while it is difficult to find the \emph{least} discriminatory alternative model for any set of equally effective models, with some effort, a model that is \emph{less} discriminatory than a baseline model can almost certainly be found in practice.

\subsection{Discovering LDAs in Practice}
\begin{figure*}[b]
    \centering
    
    \includegraphics[width=\textwidth]{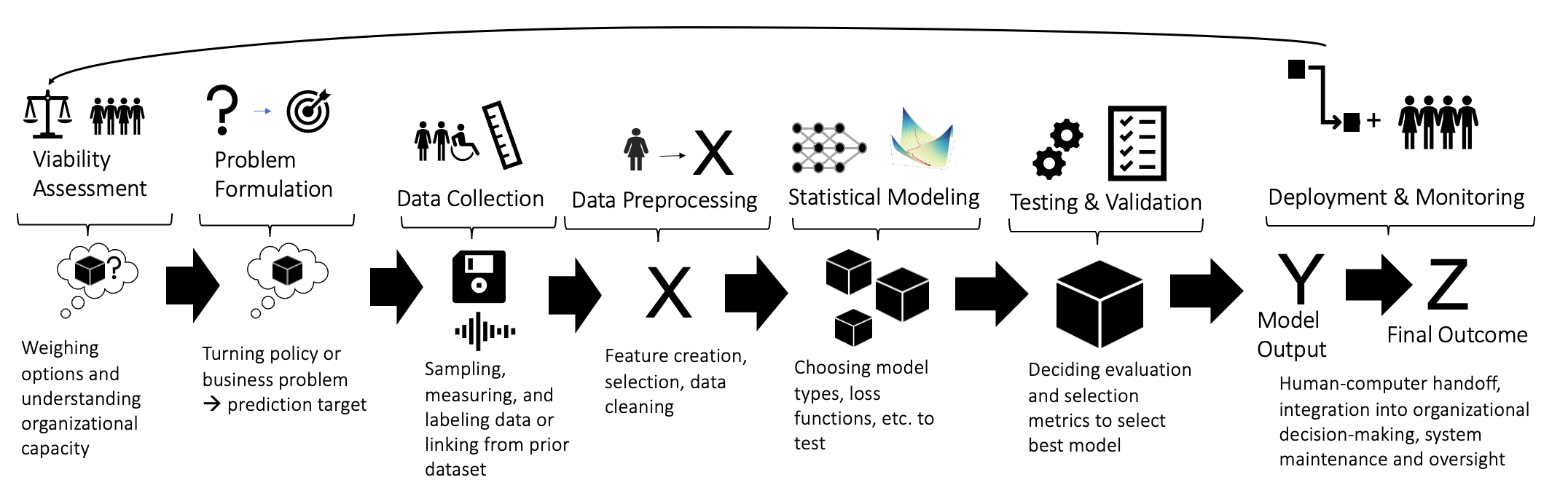}
    
    \vspace{-1.5em}
    \caption{A simplified view of the AI pipeline, its key stages, and instances of design choices made per stage.}
    \label{fig:pipeline}
\end{figure*}
Establishing evidence that model multiplicity exists, however, does not immediately explain how to find LDAs in practice. 
We suggest searching for LDAs by keeping  disparity reduction in mind throughout the model development process. As several recent works have noted, AI systems are developed through a series of iterative and subjective decisions, often called the AI development pipeline~\cite{suresh2021framework,black2023toward,coston2022validity,akpinar2022sandbox,lehr2017playing}. These decisions often involve subjective choices for which there is no correct answer, requiring the developer to weigh competing values and make judgment calls. The model development pipeline has been described by a variety of scholars so we do not fully excavate it here, but provide a brief overview in Figure~\ref{fig:pipeline}. The main stages we consider in this article are problem formulation, data collection, data preprocessing, feature selection, statistical modeling, testing and validation, and deployment and monitoring.

At every decision point, each potential choice leads to a slightly different eventual model, each with different behavior. Considering the entire machine learning pipeline, then, there is no single solution, but a range of possible models that could emerge. In order to find LDAs, model developers must create a larger suite of potential models in model development by experimenting with a wider array of design choices in the model creation pipeline: for example, experimenting with different feature sets, data imputation regimes, and hyperparameter tunings. Importantly, this experimentation regularly occurs such that developers can find models with the highest accuracy. 
To search for LDAs, practitioners can compare this wide set of models not only on the basis of accuracy, but disparity as well, and select the least discriminatory model among those that are similarly accurate. 
Not all of these potential models will have equivalent performance (i.e., performance within $\epsilon$). But as research has shown~\cite{pmlr-v139-coston21a,rodolfa2021empirical,auerbach2024testing}, many of them will. Of these equally accurate models, it is very likely that several of them will have less disparate impact.  

As we discuss in greater detail in Section ~\ref{sec:upstart}, the independent Monitorship of Upstart---a financial technology company that relies on machine learning to underwrite loans---suggests that the theoretical guarantee of model multiplicity translates into practice. As the Monitorship demonstrates, through intentional broader exploration in the model development pipeline, developers can find models with reduced disparate impact, which perform comparably to the original model. 

Though the research to date suggests that a model with reduced disparate impact can almost certainly be discovered in this manner, finding an LDA with large reductions in disparate impact is not a guarantee. There are three main reasons for this. First, there are technical limits as to the extent to which error can be redistributed to reduce disparity. For example, if the baseline model has a low error rate and a significant disparate impact, there is only so much of the disparate impact that can be reduced through exploiting model multiplicity. However, research has shown that in many circumstances relevant to civil rights law, models exhibit high error rates and high disparity, which may give developers sufficient room to redistribute error in order to meaningfully reduce disparate impact. Second, it is not clear \textit{a priori} which interventions will reduce disparities most, and thus how to search for LDAs most efficiently. However, as we discuss more in Section~\ref{sec:upstart}, the Upstart monitor's search was successful despite being relatively limited---that is, they did not explore the entire pipeline to search for LDAs---which suggests that knowing the ideal path on which to intervene is not always necessary for a successful search. Third, finding an LDA requires resources to explore interventions along the modeling pipeline. In some cases, the resources that developers would need to invest in the search process to successfully identify an LDA might exceed what they are willing or able to bear. However, as we discuss in Section ~\ref{sec:duty}, many changes in the model development process that have proven to be useful in finding LDAs are fairly trivial for most developers to integrate. Before discussing concrete methods for conducting the search for LDAs, we turn to the legal landscape regarding less discriminatory alternatives, and the implications of model multiplicity for law.

\section{Legal Background}
\label{sec:legal}
Existing civil rights laws prohibit discrimination in employment, housing, and credit. Three federal statutes are most relevant: Title VII of the Civil Rights Act of 1964 (Title VII)~\cite{title-VII} prohibits discrimination in employment; the Fair Housing Act (FHA)~\cite{fha} forbids discrimination when renting or buying a home, getting a mortgage, or seeking housing assistance; and the Equal Credit Opportunity Act (ECOA)  outlaws discriminatory financial practices~\cite{ECOA}. Each prohibits not only disparate treatment, which emphasizes discriminatory intent, but also practices that have a disparate impact. Today, disparate impact doctrine is broadly similar across these areas (though there are differences between the statutory schemes and disagreements among legal authorities interpreting the same statute). First, a plaintiff has to establish a prima facie case by showing that a policy or practice has a disparate impact on a disadvantaged group. In the second step, the defendant can avoid liability by demonstrating a legitimate business justification for the practice. Third, even if a defendant meets its burden of justifying the practice, it can still face liability if the plaintiff demonstrates that there is an alternative that would serve the same ends with less disparate impact. 

In determining how the law should take account of model multiplicity, we make two key observations. First, while many take as a given that it is a civil rights plaintiff's burden to demonstrate the existence of less discriminatory alternatives, some legal authority suggests that such alternatives are also relevant to a defendants’ burden of justification. Second, while the cases are not entirely consistent, existing authority suggests that a viable alternative does not have to be exactly equally effective as the challenged practice, and may entail some additional costs to the defendant.

\subsection{Who Bears the Burden?}
Disparate impact doctrine was first articulated in 1971 in \textit{Griggs v. Duke Power Co.}~\cite{griggs}, which involved a challenge to employment practices under Title VII. The Supreme Court held that facially neutral practices with a disparate effect on a disadvantaged minority group violated Title VII unless justified by business necessity. Immediately after \textit{Griggs}, some lower federal courts held that less discriminatory alternatives were relevant to the defendant's burden of justification. The leading case was \textit{Robinson v. Lorillard Corp.}, in which the Fourth Circuit found that proving business necessity entailed showing that there are ``no acceptable alternative policies or practices which would better accomplish the business purpose advanced, or accomplish it equally well with a lesser differential racial impact''~\cite{lorillard}.  Other courts quickly followed \textit{Lorillard's} lead.\footnote{ The Eighth Circuit adopted \textit{Lorillard’s} “no alternatives” framework in 1972, \textit{United States v. St. Louis-San Francisco Railway Co.}, 464 F.2d 301, 308 (8th Cir. 1972); the Sixth Circuit adopted this framework in 1973, \textit{Head v. Timken Roller Bearing Company}, 486 F.2d 870, 879 (6th Cir. 1973) (“The court's error arose from its failure to take into account one necessary element of the test.”); and in 1974 the Fifth Circuit held that the “nature and requirements of th[e] [business necessity] burden were correctly outlined in Robinson v. Lorillard Corp.” \textit{Pettway v. American Cast Iron Pipe Company}, 494 F.2d 211, 244-45 (5th Cir. 1974).}

In 1975, the Supreme Court signaled a different course in \textit{Albemarle Paper Company v. Moody} \cite{albemarle}. In passing, the Court described a new third step in the disparate impact framework: if an employer met its burden of proving job-relatedness, “it remains open to the complaining party to show that other tests or selection devices, without a similarly undesirable racial effect, would serve the employer’s legitimate interest.”\footnote{The Court did not discuss this possibility further, because it separately concluded that the tests were not job related and therefore unlawful.} Even though the description of this third step was dicta, the three-step process articulated in \textit{Albemarle} became the standard framework for analyzing Title VII disparate impact cases in the courts. Eventually, Congress codified disparate impact doctrine in the Civil Right Act of 1991 \cite{1991cra}, adopting the \textit{Albemarle} framework in a convoluted manner \cite{1991cra}.\footnote{For an extended discussion of this history, we refer readers to Section III of the law review edition of this paper~\cite{black2023less}.}
As amended by the Civil Rights Act of 1991 \cite{1991cra}, Title VII says that an unlawful practice is established if the plaintiff shows that a practice has a disparate impact and the defendant fails to demonstrate business necessity. Alternatively, the practice is unlawful if the plaintiff makes a demonstration “in accordance with the law as it existed on June 4, 1989, with respect to the concept of ‘alternative employment practice’”  and the defendant “refuses to adopt such alternative employment practice” \cite{title-VII}.

This odd formulation reflects a history of disagreement over the third step of the framework. On June 5, 1989, the Supreme Court decided \textit{Wards Cove v. Atonio} \cite{wardscove}, which made it significantly more difficult for plaintiffs to prevail in disparate impact cases, in part by placing a heavier burden on plaintiffs to demonstrate a less discriminatory employment practice. However, rather than clearly articulating what it means to demonstrate a less discriminatory alternative practice, Congress simply backdated the law to the day \textit{before} \textit{Wards Cove} was decided. So what was the state of less discriminatory alternatives before \textit{Wards Cove}? 

The answer depends partly on \textit{Albemarle’s} effect on \textit{Lorillard}. \textit{Albemarle} did not expressly disapprove \textit{Lorillard’s} approach of requiring defendants to show there are no alternatives with less racial impact. Rather, \textit{Albemarle} added \textit{another} route for plaintiffs to prove discrimination and did not expressly foreclose the possibility that less discriminatory alternatives remain relevant to determining whether a defendant has shown that its practice with a disparate impact is truly necessary.\footnote{We further develop this point in Section III of the law review edition of this paper~\cite{black2023less}.} 

Notably, even after \textit{Ablemarle} was decided, legal authorities continued to find the availability of alternatives to be relevant at the second step in the analysis. In 1978, the EEOC, along with three other federal agencies responsible for enforcing employment discrimination laws, issued the Uniform Guidelines on Employee Selection Procedures, indicating that an employer’s burden in justifying a test or selection procedure through a validity study includes making a reasonable search for alternatives with less adverse impact \cite{ugesp}. And even after the Civil Rights Act of 1991 explicitly placed the burden of showing an alternative employment practice on the plaintiff \cite{1991cra}, courts have still sometimes found consideration of alternatives relevant to the employer’s burden at the second step.\footnote{ For example, in \textit{EEOC v. Dial Corp.}, the court held that “[p]art of the employer’s burden to establish business necessity” includes showing that other measures “could not produce the same result.” 469 F.3d 735 (8th Cir. 2006).} 

Under the FHA, burden allocation regarding less discriminatory alternatives has also been a moving target. Like early Title VII cases, some courts found the availability of alternatives to be part of the defendant’s burden in disparate impact housing cases. In \textit{Resident Advisory Bd. v. Rizzo}, the Third Circuit held that after a plaintiff established a prima facie case, a defendant must show that its challenged practice serves “in theory and practice, a legitimate, bona fide interest” and that “no alternative course of action could be adopted that would enable the interest to be served with less discriminatory impact” \cite{rizzo}. The Second Circuit followed this reasoning~\cite{huntingtonbranch}, although other circuits placed the burden of proving a less discriminatory alternative on plaintiffs.\footnote{See, for example, \textit{Oti Kaga, Inc. v. South Dakota Housing Development Authority}, 342 F.3d 871 (8th Cir. 2003); \textit{Graoch Associates No. 33, L. P. v. Louisville/Jefferson County Metro Human Relations Commission}, 508 F.3d 366, 374 (6th Cir. 2007); \textit{Mountain Side Mobile Estates Partnership v. Secretary of Housing and Urban Development}, 56 F.3d 1243 (10th Cir. 1995).}

Agency guidance regarding who bears the burden of establishing a less discriminatory alternative in the housing context has varied over time. A 1994 policy statement by HUD and other federal agencies noted the relevance of alternatives with less discriminatory effect without indicating which party bore the burden on the issue \cite{1994fairlending}. Later, HUD indicated it would propose a rule that would “describe the standards required to demonstrate...the absence of alternatives with a less discriminatory impact,” indicating that HUD believed a defendant must show that no less discriminatory alternatives were available \cite{1994hudrule}. The following year, in a separate rulemaking, HUD suggested that if certain mortgage lenders relied on factors that have a disparate impact, they were required to demonstrate that no less discriminatory alternatives exist \cite{1995hudrule}. But HUD abandoned both efforts. Ultimately, in its 2013 disparate effects rule, HUD clarified that the plaintiff must demonstrate that “the challenged practice could be served by another practice that has a less discriminatory effect” \cite{fhacfr}, suggesting this approach “makes the most sense because it does not require either party to prove a negative'' \cite{HUDrule}.

The role of less discriminatory alternatives in the credit context has also evolved. Citing \textit{Griggs} and \textit{Albemarle}, early rulemaking efforts indicated that an “effects test” applied to creditworthiness decisions, without resolving the role of less discriminatory alternatives \cite{1977FRB}. In the 1990s, agencies enforcing ECOA and Regulation B \cite{regb} issued inconsistent guidance, at times suggesting that a creditor was required to prove a business purpose and that no less discriminatory alternative is available, and at other times backtracking from that position.\footnote{In December 1994, the Federal Reserve Board (FRB) proposed amendments to the staff commentary to Regulation B. A proposed comment regarding disparate impact and empirically derived and other credit scoring systems suggested that “credit scoring systems that employ neutral factors could violate the act or regulation if there is a disparate impact on a prohibited basis, unless the practice is justified by business necessity with no less discriminatory alternative available.” 59 Fed. Reg. 67235, 67237 (Dec. 29, 1994). In June 1995, the FRB backtracked and deleted the proposed comment. 60 Fed. Reg. 29965, 29966 (June 7, 1995). Notably the Board pointed out that ``commenters uniformly expressed concern … about the Board’s articulation of the standards of proof.'' Separately, in 1995, the National Credit Union Administration—a financial regulator with authority to enforce ECOA for federally chartered credit unions—noted that for a lender to justify a practice that has a disparate impact, ``a lender would be required to prove a business purpose for the policy and that no less discriminatory alternative is available.'' NCUA Letter to Unions, Letter No. 174, (Aug. 1995)} A 1997 bulletin from the Office of the Comptroller of the Currency on credit scoring models mentions less discriminatory alternatives, but, like the 1994 joint statement on fair lending, is silent on who bears the burden \cite{occbulletin}. Separately, interagency fair lending examination procedures suggest that plaintiffs bear the burden of demonstrating that the defendant’s legitimate business need can “reasonably be achieved as well by means that are less disparate in their impact” \cite{interagencyfairlendinappendix}.

\subsection{What is a Less Discriminatory Alternative?}

Putting aside the question of who bears the burden, a separate question remains: what exactly is a less discriminatory alternative? For years, legal scholars and practitioners have bemoaned the absence of concrete legal guidance regarding the contours of less discriminatory alternatives under Title VII, the FHA, and ECOA \cite{hsia1977, harvard1993, mahoney1998, allen2014, scherer2019, wu2024}. In parsing this question, courts and regulators have settled on a few key themes, although the details of how they apply are often contested.

First, when plaintiffs argue that a defendant should have adopted a less discriminatory alternative, they must show that such an alternative actually exists (i.e., that adopting it would have less discriminatory impact than the challenged practice)~\cite{jonesboston, citylawrence}. The alternative cannot be hypothetical or speculative---evidence must exist that the proposed alternative will actually reduce disparate impact~\cite{allenchicago, gillespie}, though what evidence is sufficient to make this determination is contested.

Second, an alternative must also advance a defendant's legitimate business purpose, though there is a great deal of disagreement about the details~\cite{kilgo, citylawrence, contreras, hardie, christner}. Uncertainty surrounds two sets of questions: first, how similar must a proposed alternative be to the defendant's current practice in advancing its goals? Must it be equally effective? Comparable? Identical? Second, to what extent are costs relevant to the determination that an alternative is a viable one? If so, which costs? 

Much of the confusion around these issues started with Title VII. As explained above, as amended by the Civil Rights Act of 1991, the statute says that one route to disparate impact liability is to make a demonstration ``in accordance with the law as it existed on June 4, 1989, with respect to the concept of ‘alternative employment practice.’”\cite{1991cra} However, the law regarding alternative employment practices was \textit{not clear} on June 4, 1989. Prior to the decision in \textit{Wards Cove} on June 5, 1989, the Supreme Court had decided only a handful of disparate impact cases, none of which offered a definitive interpretation of what an alternative practice entailed. 

In \textit{Watson v. Fort Worth Bank and Trust}~\cite{watsonbank}, decided the year before \textit{Wards Cove}, Justice O’Connor suggested that the cost or other burdens of a proposed alternative are relevant, and that a viable alternative must be equally effective; however, her views failed to command a majority. Though Justice O’Connor’s formulation regarding an alternative practice was adopted by a majority of the Court in \textit{Wards Cove}, Congress soon nullified that holding in the Civil Rights Act of 1991. While Congress failed to provide detailed guidance regarding alternative employment practices, it clearly intended to abrogate the reasoning of \textit{Wards Cove} on that issue. At a minimum then, the legislative response indicates that a viable alternative practice need not perform identically to the employer’s challenged practice and that the fact that some costs might be incurred is not decisive. 

More recently, lower courts that have considered evidence of less discriminatory alternatives in the employment context have articulated the requirements in various ways, with some holding that a proposed alternative must be ``equally effective,” or ``equally valid”~\cite{hardie, ctu}, while the EEOC has suggested a looser showing, such as ``comparably effective”~\cite{eeoc2023}.

Courts have also taken a range of approaches regarding costs. The cost of a proposed alternative remains relevant to the analysis, and when the costs of adopting a new procedure are high, it has been grounds for finding the alternative is not a viable one, particularly when it is unclear that it will produce a reduction in disparities~\cite{citylawrence}.  On the other hand, courts have found that a less discriminatory practice may entail some administrative costs. As the Court in \textit{Lorillard} explained ``some additional administrative costs may be imposed ... to eliminate discrimination, avoidance of the expense is not a business purpose that will validate an otherwise unlawful employment practice”~\cite{lorillard}.

In housing, HUD explicitly rejected an approach that would require less discriminatory alternatives to be ``equally effective,'' arguing that such a heightened standard is “less appropriate in the housing context than in the employment area in light of the wider range and variety of practices covered by the Act that are not readily quantifiable”~\cite{HUDrule}. Other regulatory guidance is consistent with this approach. In an advisory bulletin on fair lending, the Federal Housing Finance Authority (FHFA) suggested that a less discriminatory alternative need not be equally effective, but only comparably so,  and that alternatives may sometimes require entities to bear some minimal costs \cite{FHFAbulletin}.

In credit, interagency procedures have held variously that less discriminatory alternatives need to be ``approximately equally effective,” ``equally effective” \cite{interagencyfairlendinappendix}, or ``serve the same purpose with less discriminatory effect” \cite{interagencyfairlendingexam}, but provide no discussion of how costs impact an alternative’s viability. However, regulatory enforcement can offer some clues. In investigations into Honda Finance Corporation \cite{hondafinance} and Toyota Motor Credit Corporation \cite{toyotafinance}, the CFPB found that each company’s dealer markup policy was “not justified by legitimate business need and constitutes discrimination.”\footnote{Dealer markup policies allow automotive dealers to charge an interest rate above the interest rate at which a financing entity would provide financing. Under those policies, dealers would receive extra compensation from financing entities thanks to the increased interest revenue derived from the dealer markup.} In each enforcement action, the CFPB identified three potential less discriminatory alternative policies, even though each would force lenders to forego increased interest revenue. Though not formally crafted as less discriminatory alternatives, the actions suggest regulators believe that defendants may need to forego some revenue when adopting an alternative.

\section{Model Multiplicity and the Law}
\label{sec:mm_and_law}
Model multiplicity tells us that there almost always exist multiple, equally well-performing models with differing levels of disparate impact. Because of that, it makes little sense to say that a chosen model by a company covered by civil rights law is either “justified” or “necessary” if an equally accurate model that exhibits less disparate effect is available.  

As a result, in order to advance the goals of the civil rights laws, entities that use algorithms should be required to search for LDAs before deploying them in critical domains like employment, housing, and credit. The law should recognize that duty in at least two specific ways. First, under disparate impact doctrine, a defendant’s burden of justifying a model with discriminatory effects should be recognized to include showing that it made a reasonable search for LDAs before implementing it. Second, new regulatory frameworks for the governance of algorithms should include a requirement that entities undertake a search for less discriminatory models. 

\subsection{Model Multiplicity and Disparate Impact}
Under current law, after a plaintiff has established a prima facie case, the defendant bears the burden of justifying a practice with a disparate impact. It must show that the practice is “job related . . . and consistent with business necessity” under Title VII \cite{title-VII}; “necessary to achieve one or more substantial, legitimate, nondiscriminatory interests” under the FHA \cite{fhaspecific}; or “meets a legitimate business need” under ECOA \cite{regbspecific}.  Although formulated differently, the burdens placed on the defendant under each statute and their regulations make reference to need or necessity, implying that the entity cannot accomplish its goals another way. When the challenged practice involves an algorithm, it makes little sense to say that the defendant’s chosen model is “necessary” if an equally effective model that exhibits less disparate effect is available. Further, the actions required for a reasonable search in the algorithmic context are relatively unobtrusive. Developing a model through the machine learning pipeline inherently involves constant exploration and testing of alternatives. Requiring entities to also test for disparate impact and compare disparities throughout the development process is fairly easy and not terribly burdensome. 

What would civil rights litigation look like under the framework that we propose? Imagine a challenge to an algorithm used in credit, where the plaintiff alleges she was rejected for a loan because the company relied on a racially discriminatory model. The plaintiff would first have to establish a disparate impact. She might, for example, demonstrate that the model used by the lender results in disproportionately fewer Black applicants receiving loans. 

At that point, the burden would shift to the defendant to show that its practice is necessary. It would not only have to demonstrate that the model actually advances a legitimate business need and justify its definition of model performance, but also that the defendant undertook a reasonable search for LDAs before adopting the model.\footnote{This paper has discussed model performance generally. But firms will often have context-specific reasons to favor more precise definitions of performance. Even under these stricter definitions of model performance---for example, a constraint holding that models only exhibit equal performance if they get the same fraction of predictions correct and have the same false positive rate — multiplicity continues to apply. Nevertheless, each additional requirement or constraint on the definition of performance will reduce the total number of models of equivalent performance that are likely to be found. Firms may take these context-specific considerations to an extreme. In light of this risk, courts and regulators should not defer to a firm's explanation of its performance requirements, including any downstream business impact. Instead, the company should be required to justify its definition of model performance as part of its burden of showing business necessity so that viable LDAs are not arbitrarily ruled out.} A defendant might do so by producing evidence of the choices it made during the model building pipeline, such as testing the effect of changing the combination of input features on group disparities. The plaintiff, of course, might contest whether the defendant’s efforts were reasonable---for example, by arguing that the creditor failed to pursue readily available, low-cost explorations, such as comparing disparate impact of alternative models during the model development process. 

A company might have found an LDA during the development process, but decided not to adopt it. In such a case, the rejected model might be evidence that the defendant knew of a model that would have advanced its business purpose with less disparate effect, but that it refused to implement it. The defendant would bear the burden of showing that business necessity required it to implement the algorithm with greater disparate effects. Ultimately, a court would have to determine based on all the evidence whether the defendant’s efforts to search for LDAs were reasonable, and whether the availability of a viable, less discriminatory algorithm negated its claim of business necessity.

Our proposal entails no change in the third step of the disparate impact analysis. Even if a defendant satisfied its burden of showing a business justification and that its efforts to search for less discriminatory algorithms were reasonable, a plaintiff would still have the opportunity to identify a viable LDA that was overlooked by the defendant.

Practically speaking, how might our proposal be implemented? Given historic antecedents, courts applying Title VII to algorithms could simply adopt this approach in disparate impact cases involving algorithms. The EEOC could make clear in guidance that satisfying the business necessity test includes showing that the employer undertook a reasonable search for LDAs during the development process. In credit, the CFPB could formally expand upon remarks that ``[r]igorous searches for less discriminatory alternatives are a critical component of fair lending compliance management''~\cite{NCRCCFPB}.\footnote{While such public remarks from agency officials can be instructive, such remarks do not represent the formal views of the agency and are no substitute for formal guidance~\cite{Finreglab23, colfax2024report4}.} The Bureau could update its examination manuals, or update its supervisory guidance. Finally, given the specificity of HUD's existing disparate effects regulations, HUD would likely need to amend the regulation in several key places.\footnote{As one example, HUD would need to expand its definition of ``legally sufficient justification” to include reasonable efforts to search for and implement LDAs when the challenged practice is a housing algorithm.} Doing so would be directly responsive to President Biden's recent Executive Order on Safe, Secure, and Trustworthy Development and Use of Artificial Intelligence---which broadly directs agencies to use their existing authorities to combat algorithmic discrimination, and in some instances specifically directs agencies to consider or promulgate new guidance \cite{EO14110}. Using existing authorities and regulatory tools, the EEOC, HUD, and CFPB can clarify expectations regarding covered entities' duty to search for LDAs. 

\subsection{Model Multiplicity and New Regulatory Regimes}
Because of the limitations of existing civil rights laws, which rely upon a model of retrospective liability and largely depend upon individual victims of discrimination to sue to vindicate their rights, many have called for new \textit{ex ante} regulatory tools or methods \cite{cathywv2024}. A variety of interventions have been proposed—from pre-market licensing regimes~\cite{Tutt2017}, new regulatory instruments and subpoena power~\cite{EnglerCASC}, independent audits, and algorithmic impact assessments. In particular, audits and impact assessments have received significant policymaker attention. While many of these interventions are designed, at least in part, to reveal when systems lead to disparate effects on a prohibited basis, they do not offer guidance regarding what to do once an assessment or audit surfaces disparate impact. At bottom, these proposals mostly only require documentation and evaluation, and critically stop short of requiring further action. Requiring a duty to search for LDAs helps answer the questions, (1) ``What are we auditing for?” and (2) ``Algorithmic impact assessments in service of what?” 

From an anti-discrimination perspective, requiring companies to conduct reasonable searches and to adopt LDAs provides purpose and structure for those efforts. For example, when conducting an impact assessment, one critical aspect should entail assessing the choices made to explore alternative models and documenting the reasons not to pursue other paths. Similarly, if an audit reveals a disparate impact on a historically marginalized group, companies should conduct a careful search through the space of equally accurate models to ensure that a model with less discriminatory behavior is chosen. While its scope applies to federal agencies, the Office of Management and Budget’s memorandum requires federal agencies using certain rights-impacting AI systems to do exactly this: to regularly test their models for disparate impact and mitigate disparate impact when it is identified \cite{OMBAIguidance}.

Practically, new regulatory requirements could be implemented through new legislation or rulemaking under existing authority. For example, the Federal Trade Commission could promulgate a new trade rule, as part of its ongoing Commercial Surveillance and Data Security rulemaking \cite{ftcANPR2023}, requiring entities designing and deploying algorithmic systems in sensitive civil rights domains to take steps, including searching for LDAs, to address algorithmic discrimination.  An entity’s failure to do so would be an unfair trade practice, while pursuing such steps would be considered compliant with the regulation. 

With the duty’s legal scaffolding in tow, we turn to technical and practical considerations.
\section{Upstart Monitorship}
\label{sec:upstart}
Recent experience teaches that model multiplicity can be exploited in practice to actually produce viable LDAs in practical settings. Due to discrimination concerns,  Upstart (a financial technology company that relies on machine learning and non-traditional applicant data) voluntarily agreed to submit its underwriting model to scrutiny by an independent monitor,  which was tasked with assessing whether Upstart’s model had an adverse impact on any protected group, and ``if so, whether there are less discriminatory alternative practices that maintain the model’s predictiveness''~\cite{colfax2021report1}. After the Monitorship determined that Upstart's model exhibited practically and statistically significantly lower approval rates for Black applicants as compared with non-Hispanic white applicants~\cite{colfax2021report2}, the Monitorship explored whether viable LDAs existed~\cite{colfax2022report3}. To do so, the Monitor searched through every possible subset of the original model’s input features to identify combinations that yielded reductions in disparate impact and also turned to hyperparameter tuning. To evaluate potential alternatives, the Monitor relied on Upstart’s primary performance metric, which predicts both default risk and prepayment risk, and is reported as an average. Because this performance metric has some degree of uncertainty, this uncertainty can be characterized as a probability range (the ``Uncertainty Interval''). By the Monitorship’s reasoning, if the performance of an alternative model yields a value that falls within this Uncertainty Interval, there is a strong argument that it reasonably serves the same purpose as the model originally implemented by the company, since that model was already deemed acceptable to Upstart’s business purposes. As described in Section ~\ref{sec:model_mult}, identifying equally performing models requires defining some bound $\epsilon$ within which differences in performance should be considered equivalent. The Monitorship’s use of the Uncertainty Interval is an example of establishing this bound of $\epsilon$, within which differences in performance should be considered equivalent.  

Given the search process and evaluation process, the Monitor ultimately identified multiple viable LDAs:  models with lower levels of disparate impact, but with performance within a pre-determined bound $\epsilon$ of equivalent performance~\cite{colfax2022report3}. As a result, the Monitorship demonstrates that a purposeful search can result in viable LDAs---importantly, even when only a few intervention points on the ML/AI development pipeline are explored (in this case, feature selection and hyperparameter turning). However, Upstart claimed that the Monitor’s alternative ``would unacceptably compromise the accuracy of its models”\cite{colfax2024report4} and ``declined to adopt the Monitor’s recommended approach”\cite{colfax2024report4} towards finding LDAs. The parties’ disagreement centered on what the appropriate and legally required methodology is to assess whether a less discriminatory algorithm performs comparably to an existing baseline model in meeting a company’s asserted legitimate business need. As a result, the Monitorship reached an impasse.

At a practical level, however, the Monitorship highlights the shortcomings of a regime that requires civil rights plaintiffs, and not companies, to search for LDAs. The Monitorship took the better part of two years in a non-litigation, cooperative arrangement between parties to identify viable LDAs. However Upstart changed its model before an alternative could be recommended~\cite{colfax2022report3}. Had Upstart been required in the first instance to take reasonable steps to search for and implement LDAs, it’s possible, even likely, they would have discovered an LDA more quickly and at less cost and avoided unnecessary disparate impacts on real borrowers.

\section{The Duty to Search}
\label{sec:duty}
In this section, we examine the specific steps that firms should be expected to take to fulfill their duty to search for and implement LDAs in practice. 

\subsection{Basic Requirements}
Fulfilling a duty to search for and implement an LDA depends on four related processes, each of which we take to be a basic requirement of the duty. First, firms must have a process in place for collecting or inferring the demographic information necessary to perform a disparate impact analysis. 
Given the sensitivity of this data, firms must also adopt appropriate policies and procedures to protect the data and limit its use for unrelated purposes. Second, firms must have a process for actually performing the disparate impact analysis itself. Notably, this must include a process for evaluating a model's disparate impact on historically disadvantaged groups both prior to deployment and on an ongoing basis, once it has been deployed. Third, firms must establish a process for searching for LDAs. This process should apply when developing a model in the first instance, where the search for LDAs should be incorporated into the model development process from the start, and when addressing a disparate impact that has been identified after a model has been developed or deployed. A key part of this process also includes documenting at what point the firm decides to bring its search to a close---that is, why the firm believes it has done enough, given its particular constraints, to search for an LDA. Finally, firms must have a process for determining when they will adopt an LDA and for implementing the LDA in practice. 

If firms do not have these processes in place, cannot convincingly demonstrate that they have them in place, or do not disclose how they go about each process, firms should be assumed to have failed to fulfill the duty. In their absence, there is no reason to believe that firms have considered whether there is a disparate impact that might be reduced by searching for and adopting an LDA. Of course, satisfying these basic requirements alone may not satisfy the duty because the processes actually adopted may still fall short of what is reasonable. These processes could be far from robust---poorly thought through, poorly resourced, and poorly executed. Beyond these basic requirements then, we argue that firms must take reasonable steps to search for and implement LDAs. 
\subsection{Reasonable Steps}
Crucial to what counts as reasonable steps is what resources regulated entities should be expected to devote to the search for a less discriminatory alternative. As discussed in Section~\ref{sec:legal}, the law is ambiguous about how costs figure into 
determining whether a less discriminatory alternative is available. 
While courts have rejected proposed alternatives that would impose significant costs on defendants~\cite{citylawrence}, courts and regulators have also endorsed alternatives that are clearly not costless~\cite{lorillard, naacp-harrison, naacp-hudson}. These decisions suggest that there is an expectation that regulated entities incur reasonable costs in seeking to avoid disparate impact. 

In this paper, we have adopted a relatively conservative definition of an LDA—namely, it must exhibit accuracy equal to the challenged algorithm. This definition means that LDAs will not impose on firms the cost of a loss in model performance. As noted in Section~\ref{sec:legal}, this approach is more stringent than existing law, which may allow for alternatives of comparable rather than identical performance. 
By defining LDAs more stringently than legally required, we set aside questions about performance costs---the cost imposed on regulated entities if compelled to adopt an LDA of lower performance---and focus on development and administrative costs---the expense involved in finding (e.g. developing and testing a larger suite of models) and fielding 
(e.g. updating deployment procedures, such as using a model which requires more information as input) an LDA. In other words, we focus on what costs defendants should be reasonably be expected to incur in searching for and implementing LDAs.

One easy answer to these questions is that defendants should not be allowed to invoke sunk costs: the investments that they already made in deploying an algorithm that they would never have deployed in the first place if they had taken concerns with disparate impact into account from the start. 
Beyond this, at a minimum, it would be reasonable to expect firms to make any investment to reduce disparate impact that covers its own costs. For example, firms should be obligated to collect more features if doing so helps to reduce disparate impact and if the cost of doing so is covered by the additional benefits that firms enjoy from the resulting improvements in the accuracy of the decision-making process. 
Further still, firms should be expected to incur the costs of following evidence-based best practices---that is, the processes and procedures that independent research demonstrates to be effective and that experience suggests are possible for firms to execute in practice. While best practices are not a particularly precise standard, they have the benefit of being able to evolve in step with advances in our understanding of the problem and possible mitigations. They can also be responsive to norms regarding the amount of resources that industries are devoting to the problem, helping to catch out individual firms that are investing noticeably less in addressing avoidable disparate impact than their competitors. 

Finally, what counts as reasonable steps should also depend on the magnitude of the disparate impact: the more severe the disparity, the more costs firms should be expected to bear in searching for and implementing an LDA.  
\subsection{Practically Searching for LDAs}
In this section, we outline some potential techniques for searching for LDAs, while paying attention to their costs. We first note that research in this space is rapidly evolving. Methods and interventions that may be costly today may become more viable in the near future, even for companies with few available resources.  

Generally, as mentioned in Section~\ref{sec:model_mult}, the current way to find LDAs is to evaluate a suite of models created from a variety of design choices along the model development pipeline targeted at reducing disparate impact, and select the model with the least disparate impact among those with comparable accuracy.
A number of tools are available to implement disparity-aware interventions at several stages of the model development pipeline, some of which are described in a recent survey~\cite{black2023toward}, and more are likely to be available soon. Even without these tools, ongoing research provides insights on how a developer’s consideration of different choices at each step in the process might affect the degree of disparate impact exhibited by different possible models~\cite{pmlr-v139-coston21a,jeanselme2022imputation,auerbach2024testing}. A growing set of techniques can also help guide where investments should be made in the pipeline to maximize the chances of reducing disparate impact---for example, what sorts of additional data points would be most helpful in reducing disparities in accuracy rates~\cite{cai2022adaptive}. Some of these techniques can even reduce costs by suggesting interventions that are likely to have a greater return on investment than others. For example, they can suggest types of models that can update on new data without having to be fully retrained, thereby reducing costs~\cite{ren2020survey}.

While the actual cost of any given intervention will depend heavily on the particular circumstances under which a firm is operating, interventions that disrupt less of the existing pipeline are likely to be cheaper than those that involve greater change. 
For example, it would be practically trivial to add a disparity metric to 
hyperparameter tuning and doing so would help to automate the process of exploring alternatives~\cite{perrone2021fair}.
This addition would add to the cost of the computation, but these costs are likely to be modest for the relatively simple models commonly used in domains subject to discrimination law.
Generally, interventions that involve a greater disruption to the model development pipeline 
are likely to be more costly. But under certain circumstances even these can be more economical than one might expect. 
While it can be quite costly to collect the data necessary to test a range of possible target variables, it is often possible to engineer different target variables by combining or transforming existing data in different ways, thereby avoiding the cost of additional data collection. For example, models trained on IRS tax data showed reductions in disparate impact by changing the target variable from a binary outcome (is there or is there not tax fraud?) to a continuous outcome (how much money will be obtained from an audit?) without collecting new data~\cite{black2022algorithmic}.  

Almost any point of intervention can be turned into a relatively low-cost point of exploration if decisions are made in advance about what different strategies to explore. For example, while searching across all possible model types could theoretically involve testing hundreds of options, companies could define their search more narrowly, such as testing model types already in use, testing a sample of model types with similar training requirements as the baseline model, or testing model types that evidence-based best practices suggest should be used for the given application.   

Finally, high-cost interventions, such as collecting additional training data or collecting additional features 
may also improve the overall accuracy of the model and could pay for themselves. If disparate impact is reduced by improving overall accuracy instead of redistributing the same amount of error, these interventions could be not just costless, but on net profitable.  

In general, what constitutes a reasonable search depends upon the state of current research and available tools, the steps taken by other companies in an industry, the resources available to the company, and the extent of discrimination in a baseline model, among other case-by-case factors. That said, considering disparate impact at different decision points in the pipeline is a minor change from current practices to search for the most accurate models, and companies should explore the space of potential models as expansively as possible for LDAs under their particular constraints. Importantly, however, we note that the reductions in disparity must be robustly tested to ensure that they will generalize to unseen data---this can be accomplished by assessing fairness behavior with the same rigor as accuracy or other notions of performance, for example, using cross-validation or extra hold-out sets~\cite{besse2018confidence,black2024dhacking}.

\section{Caveats and Limitations}
\label{sec:caveats}
In this section, we address some of the limitations of our proposal and consider potential objections.  
\paragraph{Small Piece of Puzzle}
Ultimately, we believe that a duty to search for LDAs will help advance efforts to combat algorithmic discrimination. With reasonable efforts to search for LDAs, businesses that rely on algorithmic decision systems can avoid unnecessary disparate impacts. Of course, our proposal is just one piece of a broader puzzle. It should not be understood as the only law or policy solution to combat algorithmic discrimination. Indeed, the fact that model multiplicity can be exploited in some instances to discover LDAs should not excuse the use of fundamentally flawed algorithms that simply should not be used in the first instance. In many cases, the most effective intervention to reduce unlawful disparate impact may be for businesses to explore non-algorithmic alternatives. Nevertheless, we believe that recognizing a duty to search for LDAs is critical to fulfilling the promise of the civil rights laws.
\paragraph{Accuracy}
Our proposal defines LDAs in terms of “equal accuracy”; however, there are good reasons to be skeptical of this  approach. As a technical matter, it assumes accuracy is a stable and reliable metric. Yet there are many well-known weaknesses in the way models are commonly evaluated. For example, if incorrect labels permeate an evaluation dataset, the reported accuracy will not measure how well a model predicts the true outcome, but how well it predicts the mislabeled outcome. From a legal perspective, equal accuracy is not required; existing authorities suggests that alternatives need not be ``equally effective'' to be considered viable. We do not advocate for ``equally effective'' to be the default standard, but consider it to show that LDAs are likely available under even the \textit{most stringent definition}.
\paragraph{Students for Fair Admission}
Some may believe that the Supreme Court's recent decision in \textit{Students for Fair Admission v. Harvard} (SFFA)~\cite{SFFA}, which struck down affirmative action in college admissions as unlawful, forecloses our proposal.\footnote{For an extended discussion of \textit{SFFA} and our proposal, we refer readers to the "Limitations and Potential Objections" section of the law review edition of this paper~\cite{black2023less}.} This misapprehends our proposal and the Court's decision in \textit{SFFA}. First, the \textit{SFFA} case was decided under the Equal Protection Clause, which does not apply directly to private actors. Unlike the Equal Protection Clause, the civil rights statutes on which our argument rests already require entities to avoid unnecessary disparate impacts.

And significantly, as a descriptive matter, it is not accurate to characterize our proposal as “affirmative action.'' While the term ``affirmative action'' is sometimes used broadly to refer to any type of action taken to improve racial equity, our proposal does not entail the types of actions subjected to strict scrutiny by the Supreme Court. That doctrine aims at the use of racial \textit{classifications} to favor one group over another when making \textit{individual decisions} \cite{kim2022race, starr2024magnet}. For example, in \textit{SFFA} the Court disapproved the use of race as a ``determinative tip'' for individual applicants in the college admissions process. By contrast, when a developer examines racial impacts when choosing among models, they are not making any decisions about individual applicants that turn on their race \cite{kim2022race, starr2024magnet}. Thus, while some proposed debiasing techniques (such as training a model to equalize selection rates across groups) may raise legal concerns, our proposal does not involve the types of practices disapproved by the Court in \textit{SFFA} and other affirmative action cases. Instead, searching for LDAs entails the pursuit of equity goals through neutral means---practices that have long been legally permissible~\cite{kim2022race, starr2024magnet}.  
\section{Conclusion}
\label{sec:conclusion}
For decades, less discriminatory alternatives have been a legal backwater. Plaintiffs were poorly positioned to determine if they exist. Defendants have had little incentive to look for or implement them or they’ve resisted efforts that would have them prove they don’t exist. Model multiplicity turns the situation on its head by suggesting that in nearly all cases there are less discriminatory algorithmic alternatives to algorithms that have a disparate impact. By making reasonable efforts to search for LDAs, businesses that rely on algorithmic decision systems can avoid unnecessary disparate impacts. Recognizing a duty to make such efforts is critical to fulfilling the promise of the civil rights laws. 
\newpage
\bibliographystyle{acm}
\bibliography{bib}

\end{document}